# COVIDx-US - An Open-Access Benchmark Dataset of Ultrasound Imaging Data for AI-Driven COVID-19 Analytics

## Authors
Ashkan Ebadi[1,*], Pengcheng Xi[2], Alexander MacLean[3], Stéphane Tremblay[2], Sonny Kohli[5], & Alexander Wong[3,4]

**Affiliations**
1. National Research Council Canada, Montreal, QC H3T 1J4, Canada
2. National Research Council Canada, Ottawa, ON K1K 2E1, Canada
3. University of Waterloo, Department of Systems Design Engineering, Waterloo, ON N2L 3G1, Canada
4. Waterloo Artificial Intelligence Institute, Waterloo, ON N2L 3G1, Canada
5. Oakville Trafalgar Memorial Hospital, McMaster University, ON, Canada

[*] Corresponding author: Ashkan Ebadi (ashkan.ebadi@nrc-cnrc.gc.ca)

## Abstract
The COVID-19 pandemic continues to have a devastating effect on the health and well-being of the global population. Apart from the global health crises, the pandemic has also caused significant economic and financial difficulties and socio-physiological implications. Effective screening, triage, treatment planning, and prognostication of outcome plays a key role in controlling the pandemic. Recent studies have highlighted the role of point-of-care ultrasound imaging for COVID-19 screening and prognosis, particularly given that it is non-invasive, globally available, and easy-to-sanitize. Motivated by these attributes and the promise of artificial intelligence tools to aid clinicians, we introduce COVIDx-US, an open-access benchmark dataset of COVID-19 related ultrasound imaging data. The COVIDx-US dataset was curated from multiple sources and its current version, i.e., v1.2., consists of 150 lung ultrasound videos and 12,943 processed images of patients infected with COVID-19 infection, non-COVID-19 infection, other lung diseases/conditions, as well as normal control cases. The COVIDx-US is the largest open-access fully-curated dataset of its kind that has been systematically curated, processed, and validated specifically for the purpose of building and evaluating artificial intelligence algorithms and models.

## Background & Summary
The novel Coronavirus Disease 2019 (COVID-19), which appeared first in December 2019 and was caused by severe acute respiratory syndrome coronavirus 2 (SARS-CoV-2), led to a pandemic of severe and deadly respiratory illness, affecting human lives and well-being. The SARS-CoV-2 virus, now observed in different variants, can emerge in various forms and levels of severity, ranging from asymptomatic infection to an acute illness with organ failure risk and death[1]. Ebadi and colleagues[2] investigated the temporal evolution of COVID-19 related research themes and confirmed dynamic changes in the response of the scientific community to the disease evolution. The rapid growth of confirmed cases over several waves of a pandemic highlights the importance of effective screening and risk stratification of infected patients as a means to minimize spread and identify those that need a higher level of care[3]. The reliable and effective identification of infected patients with a low rate of false negatives contributes to controlling the disease transmission rate and mitigating the spread of the virus. A low false-positive rate is also desirable to not quarantine and treat people unnecessarily, removing burdens from the healthcare system as well as the society[4].

The reverse transcription-polymerase chain reaction (RT-PCR) test, performed on biological samples taken from the patient, is the main screening method used for COVID-19 detection[5].

Although RT-PCR is used in many countries, it requires a long complicated manual processing[3,4] that is a huge disadvantage for an effective fight against the pandemic. Moreover, there is no consensus about the sensitivity of RT-PCR testing, with highly variable rates reported in the literature[6–8]. These obstacles are compounded by a lack of necessary equipment and expertise to perform this test in many countries, an issue that also leads to improper management of infected patients[9]. Finally, RT-PCR tests do not provide additional information that supports clinical decision-making with respect to the triage of infected patients, treatment options, and predictions of patient outcomes that may assist in resource allocation. Therefore, finding complementary solutions for COVID-19 screening and alternative solutions for risk stratification and treatment planning has attracted the attention of the scientific community.

Radiography is an alternative imaging method utilized for COVID-19 screening and risk stratification. This modality entails an acute care physician and a radiologist visually inspecting radiographic images, e.g., chest X-ray (CXR) or computed tomography (CT) scans, to find indicators that are associated with SARS-CoV-2 viral infection, and that may assess the severity of infection. Biomedical imaging can accelerate diagnostic and prognostic decision-making processes by facilitating rapid assessment of patient condition and severity, as well as guiding the ordering of subsequent tests, if necessary[10]. It was reported in recent studies that patients infected with COVID-19 present abnormalities in their chest radiography images[11,12]. Additionally, some studies observed a higher sensitivity of CT scans for COVID-19 detection in their examined cohort compared to RT-PCR[7,13].

Although radiography examination is confirmed as a potential complementary method for conventional diagnostic techniques such as RT-PCR[10], some studies even suggest that it could be used as a primary COVID-19 screening tool in epidemic areas[13]. To this end, CT imaging is known to provide greater image detail and is considered as the gold standard for pneumonia detection[14]. It has also been shown to be effective for screening[7,13,15]. However, CXR imaging remains the first-line examination[10], especially in resource-limited and heavily-infected areas, mainly due to its lower cost, high availability, accessibility, and potential for rapid triaging of patients suspected of the infection[3]. Furthermore, CXR imaging has been demonstrated to be effective for both screening[3] and risk stratification[16].

As an established method for monitoring and detecting pneumonia[17], lung point-of-care ultrasound (POCUS) is an emerging imaging modality that is receiving growing attention from the scientific community in recent years[18]. Due to its many desirable properties, i.e., high portability, non-ionizing radiation nature, and being used as the preferred lung infection diagnosis and prognosis method in resource-limited settings/environments, e.g., in emergency rooms or developing countries[19], POCUS is showing considerable promise as an alternative imaging solution to CXR as the first-line screening approach[20,21], and tool that aids in prognostication[22].

Unfortunately, the literature on the applicability of POCUS for COVID-19 screening and prognosis assessment remains scarce. However, it is suggested that lung ultrasound (LUS) can play a key role in the context of the COVID-19 epidemic[10,23]. Changes in lung structure, such as pleural and interstitial thickening, are identifiable on LUS and help to detect viral pulmonary infection in the early stages[24]. For COVID-19 screening, recent studies reported identifiable lesions in the bilateral lower lobes and abnormalities in bilateral B-lines on LUS as the main attributes of the disease[25,26]. The LUS findings in other diseases, e.g., flu virus pneumonia, together with current clinical evidence, suggest that the LUS patterns of COVID-19 patients are quite characteristic, and LUS has a high potential for evaluating early lung-infected patients in various settings, including at home, patient triage, the intensive care unit, and for monitoring treatment effects[23]. Furthermore, studies have also found POCUS to be applicable for predicting mortality and whether a patient is in need of intensive care admission[22].

Artificial-intelligence (AI) powered decision support systems, mostly based on deep neural network architectures, have shown exemplary performance in many computer vision problems in healthcare[27,28]. By extracting complex hidden patterns in healthcare images, deep learning (DL) techniques may find relationships/patterns that are not instantly available to human analysis[29]. Compared to CXR and CT, lung ultrasound deep learning studies are comparably limited due to the lack of well-established, organized, carefully labelled LUS data sets[30]. Motivated by recent open-source efforts of the research community in the fight against COVID-19 and to support alternative screening, risk stratification, and treatment planning solutions powered by AI and advanced analytics, we introduce COVIDx-US, an open-access benchmark dataset of ultrasound imaging data that was carefully curated from multiple sources and integrated systematically specifically for facilitating the building and evaluation of AI-driven analytics algorithms and models. Another publicly available LUS dataset comprising of 200+ videos and ~60 images (as of April 2021 on their [GitHub repository](#)) built for COVID-19 detection is the work of Born and his colleagues[10]. As one of the main contributions of our work, in COVIDx-US we offer a systematic framework for data curation, data processing, and data validation to dataset creation for creating a unified, standardized POCUS dataset. We also tried our best to design our systematic framework to be very easy-to-use and easy-to-scale, even for users without deep computer science/programming knowledge. The current version of the COVIDx-US dataset comprises 150 videos and 12,943 processed ultrasound images of patients diagnosed with COVID-19 infection, non-COVID-19 infection, other lung diseases/conditions, as well as normal control patients. The COVIDx-US dataset was released as part of a large open-source initiative, the COVID-Net initiative[15,16,31], and will be continuously growing, as more data sources become available. To the best of the authors' knowledge, COVIDx-US is the first and largest open-access fully-curated benchmark LUS imaging dataset that is reproducible, easy-to-use, and easy-to-scale thanks to the modular well-documented design.

## Methods

The COVIDx-US dataset continues to grow as new POCUS imaging data is continuously curated and added as part of the broader initiative. All versions of the dataset will be made publicly available. Although this study represents the current snapshot of the dataset in terms of coverage, all the steps, including the data collection and processing pipeline that are introduced in this section in detail, will remain similar in the upcoming versions. Fig. **1** shows the flow of processes and the steps taken to generate the COVIDx-US dataset.

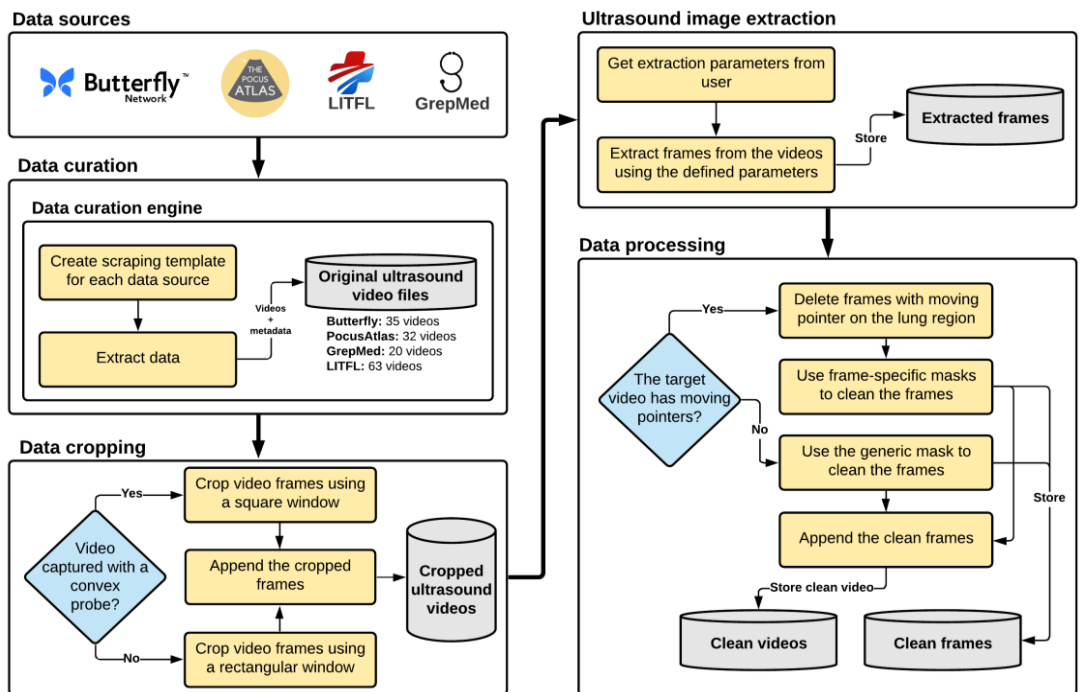

**Fig. 1** The conceptual flow of COVIDx-US data set integration. The current version of COVIDx-US contains 150 ultrasound videos and 12,943 processed ultrasound images from the following four data sources: 1) Butterfly Network, 2) GrepMed, 2) The POCUS Atlas, and 4) LITFL. Original ultrasound videos are extracted from these data sources and are curated and integrated systematically in a unified and organized structure.

## Data Sources

The COVIDx-US dataset is heterogeneous in nature, containing ultrasound imaging data of various characteristics, e.g., convex and linear US probes, from multiple sources. The current version, i.e. COVIDx-US v1.2., curates ultrasound video data of four categories, i.e., COVID-19 infection, non-COVID-19 infection (e.g., bacterial infection, non-SARS-CoV-2 viral infection, etc.), other lung diseases/conditions, and normal control, from four different sources: 1) [The POCUS Atlas](#) *(TPA)*, 2) [GrepMed](#) *(GM)*, 3) [Butterfly Network](#) *(BN)*, and 4) [Life in the Fast Lane](#) *(LITFL)*. The POCUS Atlas is a collaborative education platform for sharing ultrasound education. GrepMed is an open-access medical image and video repository. Butterfly Network is a health-tech company that developed a technology to miniaturize ultrasounds and launched a portable ultrasound device. LITFL is a repository of emergency and critical care education materials. Users are also provided with metadata to define their analytics problems as binary (COVID vs non-COVID), 3-class (COVID, non-COVID, normal), and 4-class classification problems. Table 1 shows the distribution of the LUS video files per data source in the current version of the dataset, i.e. COVIDx-US v1.2. The COVID-19 US video files account for 39% of the data, although the pandemic is recent.

**Table 1** Distribution of the collected ultrasound video files per source and class in COVIDx-US v1.2.

| Data source | Website | Categories | | | | Total |
|---|---|---|---|---|---|---|
| | | COVID-19 | Non-COVID-19 | Normal | Other | |
| TPA | www.thepocusatlas.com | 18 | 9 | 5 | 0 | 32 |
| GM | www.grepmed.com | 8 | 9 | 3 | 0 | 20 |
| BN | www.butterflynetwork.com | 33 | 0 | 2 | 0 | 35 |
| LITFL | www.litfl.com | 0 | 19 | 3 | 41 | 63 |
| | Total | 59 | 37 | 13 | 41 | 150 |

Fig. **2** shows sample ultrasound frames captured from the ultrasound video recordings in the COVIDx-US dataset. The examples are processed by the COVIDx-US scripts. These few examples illustrate the diversity of ultrasound imaging data in the dataset. The choice of the four different data sources and the heterogeneity in the structure and format of their hosted videos resulted in a highly diverse set of videos and images in the COVIDx-US dataset that is key to the generalizability of the AI-driven solutions that are built on the COVIDx-US dataset. We will continuously grow the dataset by adding more data points and/or data sources.

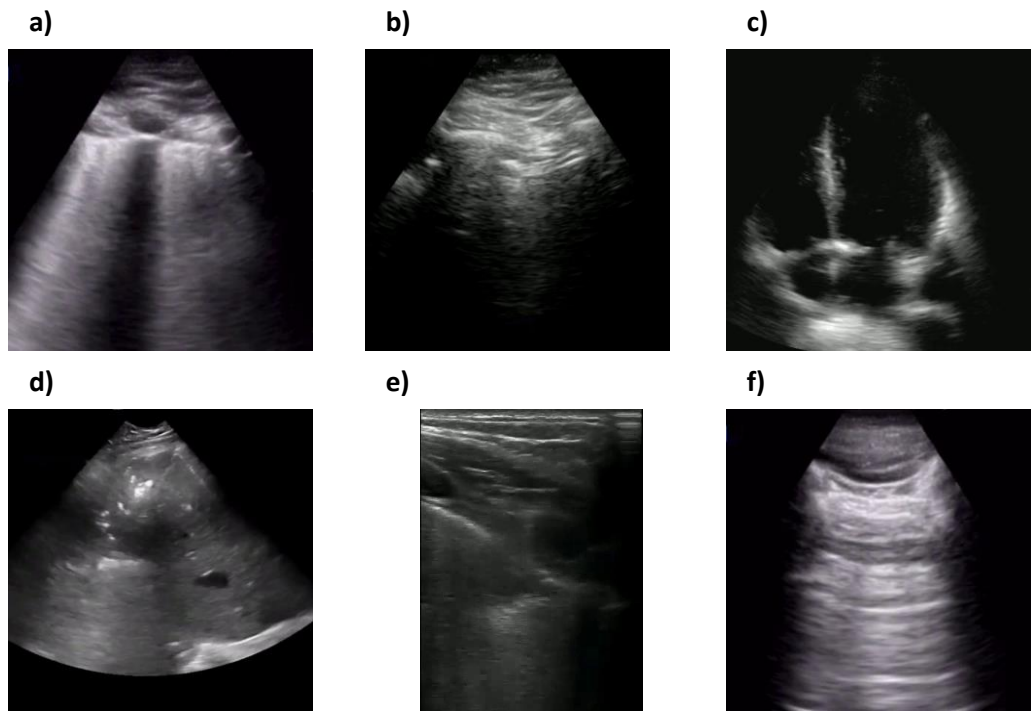

**Fig. 2** Sample ultrasound frames captured from the curated ultrasound video recordings in the COVIDx-US dataset which comprises 150 ultrasound videos, collected and curated systematically from four different data sources, and ~13,000 carefully curated ultrasound images in the current version.

## Data Curation

The data were curated from four data sources, each with a different structure. To support reproducibility and ease of use, we developed data curation engines, personalized for each of the target data sources, to automatically curate lung POCUS video recordings as well as associated metadata from the target data sources and to integrate them locally in a unified, organized structure. No original data is hosted in the COVIDx-US repository and the data is rather curated and integrated locally via our publicly released COVIDx-US scripts and the parameters set by the user. The metadata provides information on the video files, e.g., dimension and framerate, along with their category, i.e., COVID-19, non-COVID-19, other lung diseases/conditions, or normal control. The scripts are designed to be highly extensible such that more data sources can be added to the pipeline, supporting the scalability of the dataset. The scripts are made available to the general public as part of each release of the dataset.

## Data Cropping

The curated data contains video recordings captured with linear and convex US probes (N=38 and 112, respectively) that are the most common probes used in medical settings. This provides users with higher flexibility to filter in the video files based on the probe types, if required. It also enables higher generalizability of the models that are trained on the COVIDx-US dataset by covering data of different types. Fig. **3** shows examples of linear and convex US images, i.e. single snapshots of the respective video recordings. The linear probe has a flat array and appearance and provides images of higher resolution but with less tissue penetration. Convex probes, also called curved linear probes, provide a deeper and a wider view and are mostly used for abdominal scans[32]. The original data, collected from multiple sources, contains artifacts, such as measure bars, symbols, or text (Fig. **3**-a). We initially processed the collected videos and cropped them to remove these peripheral artifacts.

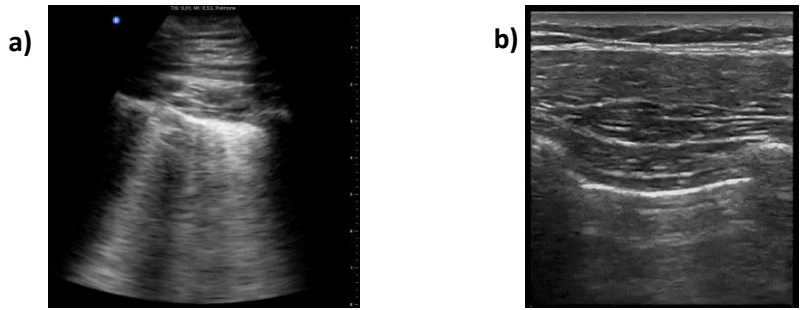

**Fig. 3** Sample frame of an ultrasound video captured with **a)** a convex, and **b)** a linear probe.

To do data cropping, we treated convex and linear US video files separately. For the convex and linear US video files, we used square and rectangular windows to crop the frames, respectively. We used rectangular windows for linear US video files to include a larger portion of the original file in the processed video file. Publicly available processing scripts that we release as part of COVIDx-US to automatically perform data cropping on the benchmark dataset. The parameters of the square and rectangular windows can be modified by the end-user, if desired. However, using the default parameters for the defined windows will remove artifacts such as bars and texts visible on the side or top of the collected US video files. The output of this step is a video file containing frames that were processed using the above-mentioned cropping process, along with a metadata file that includes information about the video file properties such as dimension and framerate, as well as the type of artifacts observed, e.g., static symbols or moving pointers. The cropped files are stored locally by the provided processing scripts.

## Ultrasound Image Extraction

As mentioned in the previous sections, the videos were curated from multiple data sources, hence, their properties differ. To ensure maximum flexibility of the COVIDx-US dataset and as part of each release, we provide end users with highly flexible data processing scripts, allowing them to extract frames from the initially processed video files based on their research objectives and requirements, using a set of parameters as follows:

- The maximum number of frames to extract from each video.
- Extract frames from either all classes or a subset of classes from the set of ['COVID-19', 'Non-COVID-19', 'Other', 'Normal'].
- Extract frames from either all data sources, i.e. ['BN, 'GM, 'LITFL', 'TPA'] or a subset of them.
- Extract frames from all videos or those captured with a specific probe, i.e. convex or linear.

We set the default parameters to extract all frames from all videos. Using the defined parameters, the frames are extracted from the videos and are stored locally.

## Data Processing

After extracting frames from the videos and using the metadata file from the data cropping stage, the frames are further processed as follows:

1. Videos with moving pointers are identified.
2. If the video contains a moving pointer:
    a. Delete frames with a moving pointer on the lung region.
    b. For the remaining frames, generate and store a frame-specific mask.
3. If the video does not contain a moving pointer:
    a. Make a generic mask (suitable for all the extracted frames) and store it.
4. Use the generated masks to process the frames, removing the remaining artifacts.

The generated masks are provided as part of the COVIDx-US release. Using the generated masks, we leveraged the inpainting technique introduced by Bertalmio and colleagues[33] to remove the remaining peripheral artifacts from the frames by replacing bad marks, i.e. pixels in the masked regions, with their neighboring pixels. The clean frames as well as the clean video file, generated by appending the clean frames, are stored locally on the user's device. Fig. **4** shows an example of a US frame, the mask generated for this specific frame, and the final clean frame obtained by applying the mask to the original frame.

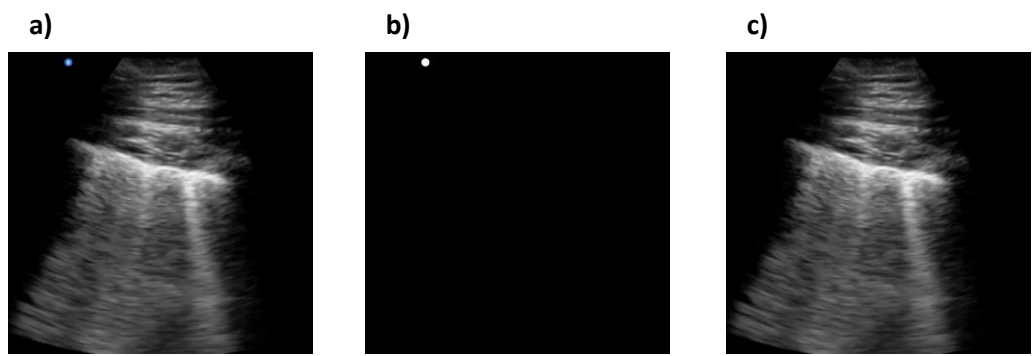

**Fig. 4 a)** A sample frame with a blue symbol on the top-left of the image, **b)** the mask generated for the frame, and **c)** the clean frame resulted from applying the generated mask to the original frame.

## Data Records

The COVIDx-US benchmark dataset is available to the general public at https://github.com/nrc-cnrc/COVID-US. The repository also includes the generated masks and metadata. The current version of the data set contains 150 processed and clean ultrasound videos, divided into 59 videos of COVID-19 infected patients, 37 videos of non-COVID-19 infected patients, 41 videos of patients with other lung diseases/conditions, and 13 videos of normal patients, along with 12,943 ultrasound images extracted from the clean video files, divided into 7,170 images of COVID-19, 3,159 images of non-COVID-19, 1,636 images of patients with other diseases/conditions, and 978 images of normal patients, using default parameters. As mentioned in the *ultrasound image extraction* section, users can extract frames from the US videos according to their projects' objectives and requirements, using the codes provided and by setting their own parameters. This makes the COVIDx-US data set highly flexible for various research objectives. Meanwhile, the modular design of the scripts allows adding/removing data sources, if required.

Running the scripts provided in COVIDx-US will extract original videos from BN, GM, TPA, and LITFL and will store them locally on the user's device in the '/data/video/original' folder. The cropped videos are stored locally in the '/data/video/cropped' folder, the clean videos in the '/data/video/clean' folder, and the clean images in the '/data/image/clean' folder. Table 2 lists video files included in COVIDx-US v1.2., and presents their properties and the number of frames extracted using the default parameters. Complementary information about the file properties can be found in the metadata files located in the '/utils' folder in the COVIDx-US GitHub repository. Users may refer to the data dictionary file located in the '/utils' folder for detailed information/description of all the metadata files. The original video files extracted from the four above-mentioned data sources are named such that the filename contains information on the source and class of the video file. This naming convention was respected for all the other generated data such as clean videos and images.

**Table 2** Ultrasound video files included in the COVIDx-US v1.2. data set.

| No | Original filename | File type | Src | Prb | Class | Original dimension | Final dimension | #Fr |
|---|---|---|---|---|---|---|---|---|
| 1 | 1_butterfly_covid | Mp4 | BN | Con | COVID-19 | 880 * 1080 | 820 * 820 | 64 |
| 2 | 2_butterfly_covid | Mp4 | BN | Con | COVID-19 | 720 * 1236 | 624 * 624 | 158 |
| 3 | 3_butterfly_covid | Mp4 | BN | Con | COVID-19 | 1928 * 1080 | 1055 * 1055 | 90 |
| 4 | 4_butterfly_covid | Mp4 | BN | Con | COVID-19 | 880 * 1080 | 820 * 820 | 108 |
| 5 | 5_butterfly_covid | Mp4 | BN | Con | COVID-19 | 860 * 1080 | 810 * 810 | 249 |
| 6 | 6_butterfly_covid | Mp4 | BN | Con | COVID-19 | 720 * 1236 | 642 * 642 | 169 |
| 7 | 7_butterfly_covid | Mp4 | BN | Con | COVID-19 | 880 * 1080 | 820 * 820 | 125 |
| 8 | 8_butterfly_covid | Mp4 | BN | Con | COVID-19 | 880 * 1080 | 820 * 820 | 109 |
| 9 | 9_butterfly_covid | Mp4 | BN | Con | COVID-19 | 1928 * 1080 | 1055 * 1055 | 80 |
| 10 | 10_butterfly_covid | Mp4 | BN | Con | COVID-19 | 736 * 1080 | 640 * 640 | 147 |
| 11 | 11_butterfly_covid | Mp4 | BN | Con | COVID-19 | 624 * 1080 | 544 * 544 | 114 |
| 12 | 12_butterfly_covid | Mp4 | BN | Con | COVID-19 | 880 * 1080 | 820 * 820 | 111 |
| 13 | 13_butterfly_covid | Mp4 | BN | Con | COVID-19 | 880 * 1080 | 820 * 820 | 91 |
| 14 | 14_butterfly_covid | Mp4 | BN | Con | COVID-19 | 880 * 1080 | 820 * 820 | 103 |
| 15 | 15_butterfly_covid | Mp4 | BN | Con | COVID-19 | 1928 * 1080 | 1055 * 1055 | 87 |
| 16 | 16_butterfly_covid | Mp4 | BN | Con | COVID-19 | 720 * 1236 | 634 * 634 | 202 |
| 17 | 17_butterfly_covid | Mp4 | BN | Con | COVID-19 | 1928 * 1080 | 1055 * 1055 | 76 |
| 18 | 18_butterfly_covid | Mp4 | BN | Con | COVID-19 | 880 * 1080 | 820 * 820 | 101 |
| 19 | 19_butterfly_covid | Mp4 | BN | Con | COVID-19 | 880 * 1080 | 820 * 820 | 81 |
| 20 | 20_butterfly_normal | Mp4 | BN | Con | Normal | 720 * 1236 | 594 * 594 | 142 |
| 21 | 21_butterfly_normal | Mp4 | BN | Con | Normal | 880 * 1080 | 820 * 820 | 99 |
| 22 | 23_grepmed_pneumonia | Mp4 | GM | Lin | Non-COVID-19 | 816 * 540 | 408 * 408 | 252 |
| 23 | 24_grepmed_covid | Mp4 | GM | Con | COVID-19 | 960 * 720 | 500 * 500 | 225 |
| 24 | 25_grepmed_pneumonia | Mp4 | GM | Con | Non-COVID-19 | 1280 * 720 | 665 * 665 | 300 |
| 25 | 26_grepmed_covid | Mp4 | GM | Con | COVID-19 | 720 * 720 | 382 * 382 | 70 |
| 26 | 27_grepmed_pneumonia | Mp4 | GM | Con | Non-COVID-19 | 480 * 360 | 345 * 345 | 91 |
| 27 | 28_grepmed_normal | Mp4 | GM | Lin | Normal | 302 * 336 | 302 * 302 | 39 |
| 28 | 29_grepmed_covid | Mp4 | GM | Lin | COVID-19 | 600 * 436 | 315 * 410 | 75 |
| 29 | 30_grepmed_covid | Mp4 | GM | Con | COVID-19 | 800 * 652 | 625 * 465 | 69 |
| 30 | 31_grepmed_covid | Mp4 | GM | Con | COVID-19 | 720 * 1076 | 608 * 608 | 365 |
| 31 | 32_grepmed_pneumonia | Mp4 | GM | Lin | Non-COVID-19 | 816 * 540 | 300 * 410 | 302 |
| 32 | 33_grepmed_covid | Mp4 | GM | Lin | COVID-19 | 960 * 720 | 435 * 500 | 116 |
| 33 | 34_grepmed_pneumonia | Mp4 | GM | Con | Non-COVID-19 | 800 * 600 | 550 * 550 | 458 |
| 34 | 35_grepmed_covid | Mp4 | GM | Con | COVID-19 | 720 * 720 | 595 * 595 | 361 |
| 35 | 36_grepmed_normal | Mp4 | GM | Con | Normal | 720 * 540 | 540 * 540 | 85 |
| 36 | 37_grepmed_pneumonia | Mp4 | GM | Con | Non-COVID-19 | 962 * 720 | 653 * 653 | 187 |
| 37 | 38_grepmed_pneumonia | Mp4 | GM | Con | Non-COVID-19 | 800 * 600 | 540 * 540 | 300 |
| 38 | 39_grepmed_normal | Mp4 | GM | Lin | Normal | 1280 * 720 | 600 * 685 | 157 |
| 39 | 40_grepmed_pneumonia | Mp4 | GM | Con | Non-COVID-19 | 500 * 354 | 354 * 354 | 114 |
| 40 | 41_grepmed_pneumonia | Mp4 | GM | Con | Non-COVID-19 | 600 * 406 | 386 * 386 | 151 |
| 41 | 42_grepmed_covid | Mp4 | GM | Con | COVID-19 | 640 * 480 | 435 * 435 | 159 |
| 42 | 43_litfl_pneumonia | Mp4 | LITFL | Con | Non-COVID-19 | 720 * 540 | 540 * 540 | 115 |
| 43 | 44_litfl_pneumonia | Mp4 | LITFL | Con | Non-COVID-19 | 720 * 540 | 540 * 540 | 18 |
| 44 | 45_litfl_pneumonia | Mp4 | LITFL | Con | Non-COVID-19 | 720 * 540 | 540 * 540 | 21 |
| 45 | 46_litfl_pneumonia | Mp4 | LITFL | Con | Non-COVID-19 | 720 * 540 | 540 * 540 | 16 |
| 46 | 47_litfl_pneumonia | Mp4 | LITFL | Lin | Non-COVID-19 | 720 * 540 | 465 * 540 | 18 |

| | | | | | | | |
|---|---|---|---|---|---|---|---|
| 47 | 48_litfl_pneumonia | Mp4 | LITFL | Con | Non-COVID-19 | 720 * 540 | 540 * 540 | 14 |
| 48 | 49_pocusatlas_covid | Gif | TPA | Con | COVID-19 | 600 * 600 | 282 * 282 | 76 |
| 49 | 50_pocusatlas_covid | Gif | TPA | Con | COVID-19 | 600 * 600 | 282 * 282 | 83 |
| 50 | 51_pocusatlas_covid | Gif | TPA | Con | COVID-19 | 600 * 1025 | 528 * 528 | 40 |
| 51 | 52_pocusatlas_covid | Gif | TPA | Con | COVID-19 | 600 * 1025 | 528 * 528 | 40 |
| 52 | 53_pocusatlas_covid | Gif | TPA | Con | COVID-19 | 598 * 430 | 400 * 320 | 41 |
| 53 | 54_pocusatlas_covid | Gif | TPA | Con | COVID-19 | 590 * 423 | 420 * 415 | 39 |
| 54 | 55_pocusatlas_covid | Gif | TPA | Lin | COVID-19 | 600 * 436 | 315 * 410 | 75 |
| 55 | 56_pocusatlas_covid | Gif | TPA | Con | COVID-19 | 600 * 410 | 410 * 410 | 30 |
| 56 | 57_pocusatlas_covid | Gif | TPA | Lin | COVID-19 | 493 * 368 | 265 * 300 | 32 |
| 57 | 58_pocusatlas_covid | Gif | TPA | Con | COVID-19 | 600 * 450 | 450 * 450 | 30 |
| 58 | 59_pocusatlas_covid | Gif | TPA | Lin | COVID-19 | 240 * 320 | 140 * 290 | 30 |
| 59 | 60_pocusatlas_covid | Gif | TPA | Con | COVID-19 | 600 * 384 | 384 * 384 | 30 |
| 60 | 61_pocusatlas_covid | Gif | TPA | Con | COVID-19 | 600 * 492 | 472 * 472 | 21 |
| 61 | 62_pocusatlas_normal | Gif | TPA | Con | Normal | 492 * 376 | 376 * 376 | 60 |
| 62 | 63_pocusatlas_covid | Gif | TPA | Lin | COVID-19 | 440 * 312 | 318 * 310 | 137 |
| 63 | 64_pocusatlas_pneumonia | Gif | TPA | Con | Non-COVID-19 | 394 * 394 | 348 * 348 | 59 |
| 64 | 65_pocusatlas_pneumonia | Gif | TPA | Lin | Non-COVID-19 | 600 * 410 | 245 * 370 | 60 |
| 65 | 66_pocusatlas_covid | Gif | TPA | Con | COVID-19 | 309 * 299 | 299 * 299 | 41 |
| 66 | 67_pocusatlas_covid | Gif | TPA | Lin | COVID-19 | 299 * 303 | 299 * 299 | 183 |
| 67 | 68_pocusatlas_pneumonia | Gif | TPA | Con | Non-COVID-19 | 282 * 290 | 282 * 282 | 30 |
| 68 | 69_pocusatlas_pneumonia | Gif | TPA | Con | Non-COVID-19 | 600 * 450 | 440 * 382 | 40 |
| 69 | 70_pocusatlas_pneumonia | Gif | TPA | Con | Non-COVID-19 | 324 * 249 | 249 * 249 | 30 |
| 70 | 71_pocusatlas_normal | Gif | TPA | Con | Normal | 600 * 450 | 450 * 450 | 59 |
| 71 | 72_pocusatlas_pneumonia | Gif | TPA | Con | Non-COVID-19 | 632 * 414 | 414 * 414 | 36 |
| 72 | 73_pocusatlas_covid | Gif | TPA | Con | COVID-19 | 439 * 595 | 407 * 407 | 46 |
| 73 | 74_pocusatlas_covid | Gif | TPA | Con | COVID-19 | 463 * 480 | 463 * 463 | 46 |
| 74 | 75_pocusatlas_pneumonia | Gif | TPA | Lin | Non-COVID-19 | 600 *409 | 285 * 350 | 61 |
| 75 | 76_pocusatlas_normal | Gif | TPA | Con | Normal | 600 * 338 | 338 * 338 | 60 |
| 76 | 77_pocusatlas_normal | Gif | TPA | Con | Normal | 237 * 293 | 237 * 237 | 60 |
| 77 | 78_pocusatlas_normal | Gif | TPA | Lin | Normal | 480 * 480 | 260 * 460 | 109 |
| 78 | 79_pocusatlas_pneumonia | Gif | TPA | Con | Non-COVID-19 | 442 * 309 | 309 * 309 | 31 |
| 79 | 80_pocusatlas_pneumonia | Gif | TPA | Con | Non-COVID-19 | 198 * 197 | 197 * 197 | 30 |
| 80 | 81_butterfly_covid | Mp4 | BN | Con | COVID-19 | 760 * 1080 | 656 * 656 | 243 |
| 81 | 82_butterfly_covid | Mp4 | BN | Con | COVID-19 | 760 * 1080 | 656 * 656 | 52 |
| 82 | 83_butterfly_covid | Mp4 | BN | Con | COVID-19 | 632 * 1080 | 558 * 558 | 76 |
| 83 | 84_butterfly_covid | Mp4 | BN | Con | COVID-19 | 624 * 1080 | 544 * 544 | 114 |
| 84 | 85_butterfly_covid | Mp4 | BN | Con | COVID-19 | 736 * 1080 | 622 * 622 | 155 |
| 85 | 86_butterfly_covid | Mp4 | BN | Con | COVID-19 | 760 * 1080 | 656 * 656 | 287 |
| 86 | 87_butterfly_covid | Mp4 | BN | Con | COVID-19 | 736 * 1080 | 640 * 640 | 177 |
| 87 | 88_butterfly_covid | Mp4 | BN | Con | COVID-19 | 760 * 1080 | 658 * 658 | 107 |
| 88 | 89_butterfly_covid | Mp4 | BN | Con | COVID-19 | 736 * 1080 | 640 * 640 | 179 |
| 89 | 90_butterfly_covid | Mp4 | BN | Con | COVID-19 | 760 * 1080 | 656 * 656 | 145 |
| 90 | 91_butterfly_covid | Mp4 | BN | Con | COVID-19 | 736 * 1080 | 640 * 640 | 402 |
| 91 | 92_butterfly_covid | Mp4 | BN | Con | COVID-19 | 736 * 1080 | 642 * 642 | 113 |
| 92 | 93_butterfly_covid | Mp4 | BN | Con | COVID-19 | 736 * 1080 | 642 * 642 | 109 |
| 93 | 94_butterfly_covid | Mp4 | BN | Con | COVID-19 | 760 * 1080 | 658 * 658 | 300 |
| 94 | 95_litfl_other | Mp4 | LITFL | Lin | Other | 480 * 360 | 360 * 310 | 45 |
| 95 | 96_litfl_other | Mp4 | LITFL | Lin | Other | 480 * 360 | 360 * 360 | 42 |
| 96 | 97_litfl_other | Mp4 | LITFL | Con | Other | 480 * 360 | 360 * 360 | 29 |
| 97 | 98_litfl_other | Mp4 | LITFL | Con | Other | 480 * 360 | 360 * 360 | 149 |
| 98 | 99_litfl_other | Mp4 | LITFL | Lin | Other | 480 * 360 | 335 * 472 | 46 |
| 99 | 100_litfl_other | Mp4 | LITFL | Con | Other | 480 * 360 | 360 * 360 | 46 |
| 100 | 101_litfl_other | Mp4 | LITFL | Con | Other | 480 * 360 | 360 * 360 | 28 |
| 101 | 102_litfl_other | Mp4 | LITFL | Con | Other | 480 * 360 | 360 * 360 | 28 |
| 102 | 103_litfl_other | Mp4 | LITFL | Lin | Other | 480 * 360 | 335 * 462 | 46 |
| 103 | 104_litfl_other | Mp4 | LITFL | Lin | Other | 480 * 360 | 340 * 463 | 46 |
| 104 | 105_litfl_other | Mp4 | LITFL | Con | Other | 480 * 360 | 360 * 360 | 43 |
| 105 | 106_litfl_other | Mp4 | LITFL | Lin | Other | 480 * 360 | 360 * 380 | 39 |
| 106 | 107_litfl_other | Mp4 | LITFL | Con | Other | 480 * 360 | 360 * 360 | 43 |
| 107 | 108_litfl_other | Mp4 | LITFL | Con | Other | 480 * 360 | 360 * 360 | 43 |
| 108 | 109_litfl_other | Mp4 | LITFL | Lin | Other | 480 * 360 | 360 * 480 | 31 |
| 109 | 110_litfl_other | Mp4 | LITFL | Lin | Other | 480 * 360 | 350 * 468 | 29 |
| 110 | 111_litfl_other | Mp4 | LITFL | Con | Other | 480 * 360 | 360 * 360 | 33 |
| 111 | 112_litfl_other | Mp4 | LITFL | Lin | Other | 480 * 360 | 355 * 470 | 37 |
| 112 | 113_litfl_other | Mp4 | LITFL | Con | Other | 480 * 360 | 360 * 360 | 45 |
| 113 | 114_litfl_other | Mp4 | LITFL | Con | Other | 480 * 360 | 360 * 360 | 46 |
| 114 | 115_litfl_other | Mp4 | LITFL | Con | Other | 480 * 360 | 360 * 360 | 28 |
| 115 | 116_litfl_other | Mp4 | LITFL | Con | Other | 480 * 360 | 360 * 360 | 28 |
| 116 | 117_litfl_other | Mp4 | LITFL | Con | Other | 480 * 360 | 360 * 360 | 39 |
| 117 | 118_litfl_other | Mp4 | LITFL | Con | Other | 480 * 360 | 360 * 360 | 28 |
| 118 | 119_litfl_other | Mp4 | LITFL | Con | Other | 480 * 360 | 360 * 360 | 28 |
| 119 | 120_litfl_other | Mp4 | LITFL | Con | Other | 480 * 360 | 360 * 360 | 28 |
| 120 | 121_litfl_pneumonia | Mp4 | LITFL | Con | Non-COVID-19 | 480 * 360 | 360 * 360 | 27 |
| 121 | 122_litfl_pneumonia | Mp4 | LITFL | Con | Non-COVID-19 | 480 * 360 | 360 * 360 | 42 |
| 122 | 123_litfl_pneumonia | Mp4 | LITFL | Con | Non-COVID-19 | 480 * 360 | 360 * 360 | 36 |
| 123 | 124_litfl_pneumonia | Mp4 | LITFL | Lin | Non-COVID-19 | 480 * 360 | 355 * 420 | 36 |
| 124 | 125_litfl_pneumonia | Mp4 | LITFL | Lin | Non-COVID-19 | 480 * 360 | 360 * 460 | 36 |
| 125 | 126_litfl_pneumonia | Mp4 | LITFL | Lin | Non-COVID-19 | 480 * 360 | 360 * 460 | 31 |
| 126 | 127_litfl_pneumonia | Mp4 | LITFL | Con | Non-COVID-19 | 480 * 360 | 360 * 360 | 24 |
| 127 | 128_litfl_pneumonia | Mp4 | LITFL | Con | Non-COVID-19 | 480 * 360 | 360 * 360 | 32 |
| 128 | 129_litfl_pneumonia | Mp4 | LITFL | Con | Non-COVID-19 | 480 * 360 | 360 * 360 | 25 |
| 129 | 130_litfl_pneumonia | Mp4 | LITFL | Con | Non-COVID-19 | 480 * 360 | 360 * 360 | 41 |
| 130 | 131_litfl_pneumonia | Mp4 | LITFL | Con | Non-COVID-19 | 480 * 360 | 360 * 360 | 26 |
| 131 | 132_litfl_pneumonia | Mp4 | LITFL | Con | Non-COVID-19 | 480 * 360 | 360 * 360 | 27 |
| 132 | 133_litfl_other | Mp4 | LITFL | Lin | Other | 480 * 360 | 360 * 410 | 46 |
| 133 | 134_litfl_other | Mp4 | LITFL | Lin | Other | 480 * 360 | 355 * 400 | 46 |
| 134 | 135_litfl_normal | Mp4 | LITFL | Lin | Normal | 480 * 360 | 360 * 410 | 36 |
| 135 | 136_litfl_other | Mp4 | LITFL | Lin | Other | 480 * 360 | 360 * 410 | 36 |
| 136 | 137_litfl_other | Mp4 | LITFL | Lin | Other | 480 * 360 | 330 * 470 | 46 |
| 137 | 138_litfl_normal | Mp4 | LITFL | Lin | Normal | 480 * 360 | 360 * 460 | 46 |
| 138 | 139_litfl_normal | Mp4 | LITFL | Lin | Normal | 480 * 384 | 384 * 430 | 26 |
| 139 | 140_litfl_other | Mp4 | LITFL | Lin | Other | 480 * 384 | 365 * 428 | 27 |
| 140 | 141_litfl_other | Mp4 | LITFL | Con | Other | 480 * 384 | 384 * 384 | 46 |
| 141 | 142_litfl_other | Mp4 | LITFL | Lin | Other | 480 * 360 | 340 * 430 | 34 |
| 142 | 143_litfl_other | Mp4 | LITFL | Lin | Other | 480 * 360 | 360 * 405 | 38 |
| 143 | 144_litfl_other | Mp4 | LITFL | Lin | Other | 480 * 360 | 360 * 405 | 46 |
| 144 | 145_litfl_other | Mp4 | LITFL | Con | Other | 480 * 360 | 360 * 360 | 45 |
| 145 | 146_litfl_other | Mp4 | LITFL | Con | Other | 480 * 360 | 360 * 360 | 45 |
| 146 | 147_litfl_other | Mp4 | LITFL | Con | Other | 480 * 360 | 360 * 360 | 36 |
| 147 | 148_litfl_other | Mp4 | LITFL | Con | Other | 480 * 360 | 360 * 360 | 26 |
| 148 | 149_litfl_other | Mp4 | LITFL | Con | Other | 480 * 360 | 360 * 360 | 18 |
| 149 | 150_litfl_other | Mp4 | LITFL | Con | Other | 480 * 360 | 360 * 360 | 28 |
| 150 | 151_litfl_pneumonia | Mp4 | LITFL | Con | Non-COVID-19 | 480 * 360 | 360 * 360 | 42 |
| | | | | | | | **Total** | 12943 |

**Note: Src:** Data source, **Prb:** probe type, **#Fr:** Number of frames.

## Technical Validation

The COVIDx-US is curated from four data sources and contains data of different types and characteristics. The scripts provided will perform the processes necessary to clean the collected POCUS videos, extract frames, and store them locally on the user's device. But, they do not validate the analyses performed and published by the research community using COVIDx-US data. As COVIDx-US will be continuously growing, feedbacks provided by researchers will provide information that may be used in the next versions of COVIDx-US to perform additional processes/reviews. Such feedbacks may be addressed to ashkan.ebadi@nrc-cnrc.gc.ca.

In order to validate the quality of images in the COVIDx-US dataset and ensure the existence of markers in the processed ultrasound images, our contributing clinician (S.K.) reviewed a randomly selected set of images and reported his findings and observations. Our contributing clinician is a practicing Internal Medicine and ICU (Intensive Care specialist), certified in both specialties by the Royal College of Physicians of Canada. Fig. **5** shows three select images of COVID-19 positive cases, as examples, that were reviewed. The summary of our expert clinician's report is as follows.

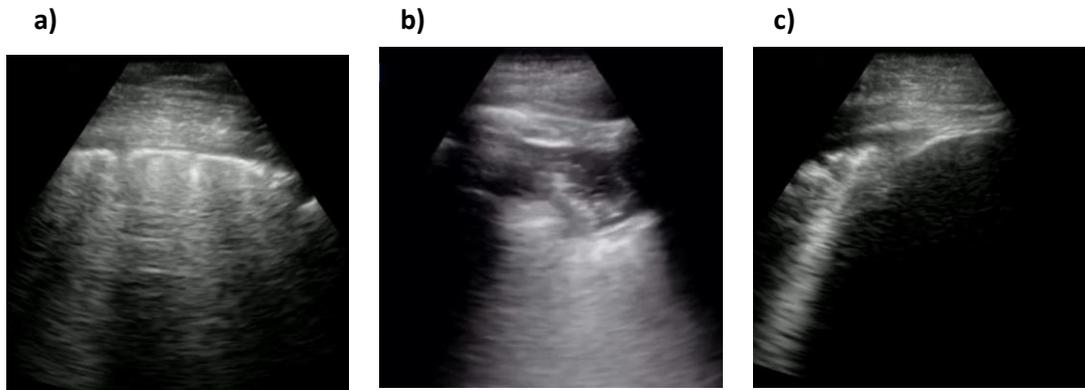

**Fig. 5** Sample processed ultrasound images of COVID-19 positive cases, reviewed and reported on by our contributing clinician.

**Case 1 (Fig. 5-a).** Our contributing clinician observed multiple pleural irregularities, including pleural thickening and the presence of sub-pleural consolidations which have been previously described as markers of COVID-19 disease severity[34]. These findings, together with the observed C-line profile, are indicative of a moderate to severe pulmonary disease.

**Case 2 (Fig. 5-b).** This is an image of lung pleura in short depth. Our clinical expert observed abnormalities and irregularities in the pleura as it is thickened and "shredded" with hypoechoic signals suggesting consolidations and air bronchograms. Although a deeper view to assess for B-lines would be more optimal, these findings together suggest moderate airspace disease, most commonly on the basis pneumonia.

**Case 3 (Fig. 5-c).** This appears to be an image of lung pleura at depth of ~5±2cm. According to our contributing clinician, the pleura and underlying lung are abnormal. There is the presence of a "waterfall sign", i.e., subpleural consolidation with a B-line. The pleura is thickened and irregular. Despite the observed abnormalities, more imaging data is needed to inform on differential diagnosis. In addition, it is not possible to comment on lung sliding as this is a static image.

Our expert clinician findings and observations confirmed the existence of identifiers and indicators of disease in the COVIDx-US dataset. AI-powered analytics solutions can exploit such indicators and patterns to detect COVID-19. Based on our contributing clinician's evolving experience, LUS has significant utility in the management of COVID-19 patients with respiratory symptoms. As a safe, rapid, reproducible, low-cost, and highly informative tool for assessing the severity of lung involvement, early studies suggest that it can be used to inform triage and treatment decisions[35]. To this end, several published LUS-based protocols are now undergoing validation in prospective clinical trials[36,37]. Furthermore, several groups are now evaluating the potential utility of LUS in other settings, including the Intensive Care Unit (ICU) where it could be used to track disease progression, and to evaluate patient candidacy and clinical response to various interventions including ventilator weaning, prone positioning and lung recruitment maneuvers in patients with acute respiratory distress syndrome (ARDS)[38].

Known limitations of this modality include the observation that LUS findings are not necessarily specific to COVID-19. Moreover, they have yet to be proven as reliable markers of clinical outcome in appropriately sized clinical studies. The deployment of LUS in COVID-19 also requires strict infection control measures. Lastly, LUS requires significant operator training and experience before it can be used in the management of potentially unstable patients, or in those with suspected infectious syndromes.' AI-driven solutions can aid clinicians with the screening process of COVID-19 patients, reducing the pressure on healthcare systems and healthcare providers.

## Usage Notes

We are constantly searching for more data, therefore, the COVIDx-US will be growing over time as more data sources become available. We recommend that users check the COVIDx-US repository at https://github.com/nrc-cnrc/COVID-US, for the latest version of data and scripts. The data collection and processing pipeline is coded in Python (version 3.6.12). Users are provided with a Python notebook including all the steps required to collect, process, and integrate data, as described in the manuscript. The provided scripts are well-documented allowing users to modify parameters for frame extraction from ultrasound videos, based on their research objectives and requirements, if required.

## Code Availability

All the codes and materials, e.g. metadata and masks, necessary to reproduce the COVIDx-US data set, as described and explained in this manuscript, are available to the general public at https://github.com/nrc-cnrc/COVID-US, accessible with no restrictions. The scripts were in Python programming language (version 3.6.12), using pandas 1.1.3, selenium 3.141.0, and requests 2.24.0 libraries.

## Acknowledgements

The authors would like to offer their special thanks to Mr. Patrick Paul for his help and support during this project.